\documentclass[final]{siamltex}
\usepackage{amsmath,bm}

\usepackage{amsfonts}
\usepackage{graphicx}

\usepackage{xparse}
\usepackage{soul}
\usepackage{xcolor}

\usepackage{algorithm}
\usepackage{algpseudocode}

\makeatletter
\NewDocumentCommand{\sotwo}{O{red}O{black}+m}
    {%
        \begingroup
        \setulcolor{#1}%
        \setul{-.5ex}{.4pt}%
        \def\SOUL@uleverysyllable{%
            \rlap{%
                \color{#2}\the\SOUL@syllable
                \SOUL@setkern\SOUL@charkern}%
            \SOUL@ulunderline{%
                \phantom{\the\SOUL@syllable}}%
        }%
        \ul{#3}%
        \endgroup
    }
\makeatother

\DeclareMathAlphabet\mathpzc{OT1}{pzc}{m}{it}
\let\mathcal=\mathpzc

\let\trueiiint=\iiint
\def\iiint{\mathop{\textstyle\trueiiint}\limits}
\def\intinfty{\int\limits_{\!\!-\infty\,\,}^{\,\,\infty\!\!}\kern-0.0em}
\def\iintinfty{\mathop{\int\!\!\int}\limits_{\!\!-\infty\,\,}^{\,\,\infty\!\!}\kern-0.0em}
\def\iiintinfty{\mathop{\int\!\!\int\!\!\int}\limits_{\!\!-\infty\,\,}^{\,\,\infty\!\!}\kern-0.0em}

\let\<=\langle
\let\>=\rangle
\let\^=\hat
\def\~#1{{\-ox{\sf#1}}}

\let\-=\mathbf

\def\circ{\ifmmode\mathchar"220E\else$\mathchar"220E$\fi}
\def\@#1{{\cal #1}}

\title{Maximum  conditional entropy Hamiltonian Monte Carlo sampler\thanks{The work was  supported by the NSFC, under grant number 11301337.}}

\author{Tengchao Yu\footnotemark[2]
\and Hongqiao Wang\footnotemark[3]
        \and Jinglai Li\footnotemark[4]}

\begin{document}

\maketitle

\renewcommand{\thefootnote}{\fnsymbol{footnote}}

\footnotetext[2]{School of Mathematical Sciences,  
Shanghai Jiao Tong University,  Shanghai 200240, China, (tengchaoyu@sjtu.edu.cn).}
\footnotetext[3]{School of Mathematics and Statistics, Central South University, Changsha, Hunan 410083, China, 
(wanghongqiao@csu.edu.cn).}
\footnotetext[4]{Corresponding Author, School of Mathematics, University of Birmingham, 
Birmingham B15 2TT, UK, (j.li.10@bham.ac.uk).}

\renewcommand{\thefootnote}{\arabic{footnote}}

\begin{abstract}
The performance of Hamiltonian Monte Carlo (HMC) sampler depends
critically on some algorithm parameters such as the total integration time
and the numerical integration stepsize. 
The parameter tuning is particularly challenging when the mass matrix of the HMC sampler is adapted. 
We propose in this work a Kolmogorov-Sinai entropy (KSE) based design criterion to optimize these
algorithm parameters,
which can avoid some potential issues in the often used jumping-distance based measures.
For near-Gaussian distributions, we are able to derive the optimal algorithm parameters with respect 
to the KSE criterion analytically. 
As a byproduct the KSE criterion also provides a theoretical justification 
for the need to adapt the mass matrix in HMC sampler. 
Based on the results, we propose 
an adaptive HMC algorithm, and we then demonstrate the performance of the proposed algorithm with numerical examples. 
\end{abstract}

\pagestyle{myheadings}
\thispagestyle{plain}
\markboth{T. YU, H. WANG AND J. LI}{ADAPTIVE MCMC FOR BAYESIAN INFERENCE}

\section{Introduction}

Generating samples from a target distribution is an important practice in many fields of science and engineering. 
The Hamiltonian Monte Carlo (HMC) method~\cite{neal2011mcmc,barp2018geometry}, initially developed in the Physics community~\cite{duane1987hybrid}, has become a very popular tool 
for sampling distributions with continuously differentiable density functions. 
The HMC method receives considerable attention from Scientific Computing and Computational Statistics, and many significant improvements of the method have been been developed in the last decade, for example,
~\cite{bui2014solving,chen2014stochastic, girolami2011riemann, hoffman2014no,strathmann2015gradient,wang2013adaptive,zhang2011quasi}, 
just to name a few. 
Loosely speaking, the HMC method proposes new samples by simulating particle movement of a Hamiltonian system
constructed from the target density for a given amount of time, and it can often achieve better performance than 
standard random-walk based Markov chain Monte Carlo (MCMC) algorithms as it takes advantage of the gradient information of the target distribution~\cite{neal2011mcmc,barp2018geometry}.

It is well-known that the performance of the HMC algorithm depends critically on the algorithm parameters, and 
in particular on  the mass matrix or the metric~\cite{betancourt2017conceptual}, 
the total integration time and the numerical integration stepsize.  
As such, tuning these parameters becomes an important task in the implementation of HMC.
First, a popular practice to tune the mass matrix is to set it to be the inverse of the covariance of the target distribution,
and the physical intuition behind this choice can be found in e.g., \cite{neal2011mcmc}. 
A rigorous justification of such a choice has yet been established to the best of our knowledge. 
On the other hand,  a number of methods have been developed to tune the other two key algorithm parameters:
the total integration time and the integration stepsize. 
Such works include, 
the optimal tuning derived in the infinite dimensional limit~\cite{beskos2013optimal}, 
the No U-turn sampler (NUTS)~\cite{hoffman2014no}, and the Adaptive HMC sampler (AHMC)~\cite{wang2013adaptive}.
However, it is rather difficult to directly apply these methods when the mass matrix is adapted during the MCMC 
iteration: for example, in the implementation of NUTS in STAN~\cite{carpenter2017stan} it is suggested that the mass matrix is estimated in the warmup period and fixed afterwards.   
The main problem concerned in this work is how to determine the integration time and the numerical stepsize
when the mass matrix is adapted. 

When designing a tuning scheme of MCMC, one usually relies on a specific design criterion or performance measure,
and for example, in both~\cite{beskos2013optimal, wang2013adaptive}, the tuning schemes make use of the expected squared jumping distance (ESJD), which was proposed in ~\cite{Pasarica2007Adaptively} for random walk MCMC algorithms.   
The ESJD criterion, which seeks the largest jumping distance,  may cause some potential issues in the HMC sampler. 
An intuitive example for this 
is that the particles jumps from one side of the space to the other in each iteration, and in this case even though they may travel
a very long distance, the Markov chain may not even be ergodic~\cite{neal2011mcmc}.
In Section~\ref{sec:gaussian} we shall demonstrate that this is exactly the case when the HMC proposal is tuned by maximizing the ESJD for the simple Gaussian target distributions. To address the issue we propose a new design criterion based on the 
Kolmogorov-Sinai entropy (KSE)~\cite{sinai2009kolmogorov} of the underlying Markov chain and we show that this criterion can lead to the very optimal proposal in the Gaussian case (namely, when the target distribution is exactly Gaussian, the proposal becomes the target distribution itself).
Moreover, for near-Gaussian distributions, we can analytically derive the optimal integration time with respect to 
the KSE (which, as will be shown in Section~\ref{sec:mces}, is essentially the conditional entropy 
of the underlying Markov chain) criterion, and that enables us to develop a HMC algorithm that adapts the mass matrix. 
Interestingly we are also able to show that the optimal mass matrix under the KSE criterion is 
actually the inverse of the target covariance,
providing a justification of the strategy previously obtained from physical intuition~\cite{neal2011mcmc}. 
With numerical examples, we show that the proposed algorithm has a rather competitive performance 
against existing methods for a certain class of problems, i.e., those with near-Gaussian distributions. 

To summarize, the \emph{main contribution} of the present work is two fold: first we provide a KSE based design criterion 
to determine the algorithm parameters of HMC proposal, and for near-Gaussian distributions, we derive 
an analytical result of the optimal integration time under the KSE criterion;
second, using the optimality results we design an adaptive HMC algorithm which automatically determines the stepsize
while the mass matrix is adapted. 
The rest of the paper is organized as follows.
In Section~\ref{sec:mces} we present the proposed Maximum conditional entropy (or KSE) HMC sampler,
and provide some optimality results of the sampler.
In Section~\ref{sec:examples} we demonstrate the performance of the proposed algorithm with 
mathematical and practical examples. 
Finally in Section~\ref{sec:conclusions}, we provide some concluding remarks
on both the advantages and the limitations of the proposed method.

\section{The maximum conditional entropy HMC sampler} \label{sec:mces}
\subsection{The HMC algorithm}
Let $\-x$ be a $n$-dimensional random variable with distribution 
\begin{equation}
\pi(\-x) =\exp(-U(\-x)).
\end{equation}
By using the Hamiltonian dynamics, one can design a very efficient scheme to draw samples from the distribution $\pi(\-x)$. 
Specifically one constructs an artificial Hamiltonian system
that has $\-x$ as its position variable and the function $U(\-x)$ as the potential energy of the system,
and then introduces an auxiliary variable $\-p$ to be the momentum of the system with kinetic energy $K(\-p)$.
In practice, the kinetic energy is usually taken to be of a quadratic form:
\begin{equation}
K(\-p) = \frac12\-p^T M^{-1}\-p,\label{e:kp}
\end{equation} 
where $M$ is a positive definite symmetric matrix. 
The dynamics of the constructed system is governed by
\begin{equation}
\label{eq:hamiltonian}
\frac{d\-x}{dt}=\frac{\partial H}{\partial\-p},\quad\frac{d\-{p}}{dt}=-\frac{\partial H}{\partial \-{x}}.
\end{equation}
 Suppose the current position is $\-x_0$ and the HMC performs the following steps in each iteration, 
\begin{itemize}
  \item Sample an initial momentum state $\-p_0$ from the distribution $N(0,M)$;
  \item Solve the Eq.~\eqref{eq:hamiltonian} with initial condition $(\-x_0,\-p_0)$, for a given amount of time $T$, 
	obtaining the new states $(\-x_T,\-p_T)$;
	\item Accept the new position $\-x_T$  with probability
      \begin{equation}
      \min [1,\exp(H(\-x_0,\-p_0)-H(\-x_T,\-p_T))]. \label{e:accprob}
      \end{equation}
\end{itemize}
Since the Hamiltonian system preserves its total energy, 
\[H(\-x(t),\-p(t))=U(\-x(t))+K(\-p(t)),\] 
it should be clear that if we can solve the Eq.~\eqref{eq:hamiltonian} exactly, 
the acceptance probability is simply one. In practice, however, Eq.~\eqref{eq:hamiltonian} must be solved numerically
and as a result the acceptance probability is lower than  one.
The leapfrog algorithm~\cite{neal2011mcmc}  is commonly used to solve the system for its ability to preserve the time-reversibility,
and moreover, when the integration time $T$ is large, one often use multiple leapfrog steps to integrate the ODE system~\eqref{eq:hamiltonian} from $0$ to $T$. In Alg.~\ref{alg:leapfrog} we present the multiple-step leapfrog algorithm for solving the ODE system~\eqref{eq:hamiltonian}. 
\begin{algorithm}[!t]
\let\-=\mathbf
\begin{algorithmic}[1]
\Function{$(\-x_T,\-p_T)$={leapfrog}}{$\mathbf{x}_0, \mathbf{p}_0, M, U,\epsilon, L$}
\For {$i = 1$ to $L$}
	\State Set $\mathbf{p} \leftarrow \mathbf{p}_0+\frac{1}{2}\epsilon\nabla U(\mathbf{x}_0)$;
	\State Set $\mathbf{x} \leftarrow \mathbf{x}_0+\epsilon{M}^{-1}\mathbf{p}$
	\State Set $\mathbf{p} \leftarrow \mathbf{p}+\frac{1}{2}\epsilon\nabla U(\mathbf{x})$;  
\EndFor 
\State Set $\-x_T \leftarrow \-x$ and $\-p_T \leftarrow \-p$; 
\EndFunction
\end{algorithmic}
\caption{The leapfrog algorithm}\label{alg:leapfrog}
\end{algorithm}

\subsection{Maximizing the entropy rate}\label{sec:mce}
Now suppose that we can solve the Hamiltonian system~\eqref{eq:hamiltonian} exactly, achieving the $100\%$ acceptance probability, 
and an important question here is \textit{how to choose matrix $M$ and time $T$ for the best efficiency of the HMC algorithm?}
To answer this question, we first need to establish an optimality criterion or a performance measure of the HMC sampler. 

A very natural choice for such a performance measure is the ESJD:
\begin{equation}
\mathbb{E}[\|\-x_T-\-x_0\|^2], \label{e:esjd}
\end{equation}
 which was first proposed by Pasarica and Gelman~\cite{Pasarica2007Adaptively} 
and was later applied to HMC in \cite{beskos2013optimal, wang2013adaptive}.  
As is mentioned earlier the main idea behind the ESJD criterion is to seek a proposal that moves the largest distance from the present location, 
and this idea performs well in a number of target distributions. 
However, as will be shown later, maximizing the ESJD may lead to problematic HMC proposals in certain circumstances
and in particular, in the simple Gaussian case, the resulting chain becomes periodic and loses its ergodicity. 
Naturally we expect that the ESJD criterion may not perform well for distributions that are close to Gaussian, 
an important class of distributions in many practical problems. 
This issue then motivates us to consider alternative design criterion. 
It is well known that a general principle to design a MCMC scheme
is to minimize its mixing time, which however can not be directly used 
as it is extremely difficult to compute. 
However, it is reported in \cite{mihelich2018maximum} 
that maximizing the KSE can yield very close results 
to those of minimizing the mixing time and for general Markov chains.
Based on this finding we propose to determine the parameters in HMC by maximizing the KSE.
For a stationary time-reversible Markov chain, the KSE is equal to the conditional entropy~(CE): 
\begin{equation}\label{e:ce}
\mathbb{H}(\-x_T|\-x_0) =\int \log \pi(\-x_T|\-x_0) \pi(\-x_T,\-x_0)d\-x_Td\-x_0,
\end{equation}
which means that the algorithm parameters are determined by maximizing the CE of the chain. 
When $\pi(\-x_T|\-x_0)$ is Gaussian, the CE in Eq.~\eqref{e:ce} is (up to a constant) equal to 
\begin{equation}
\mathbb{E}_{\-x_0}[\log\det\mathrm{Cov}[\-x_T|\-x_0]]. \label{e:convar}
\end{equation} 
Thus when $\pi(\-x_T|\-x_0)$ is Gaussian or near-Gaussian, we can choose to optimize Eq.~\eqref{e:convar} for that it is usually easier to evaluate than Eq.~\eqref{e:ce}.


\subsection{Near-Gaussian distributions}\label{sec:gaussian}
In this section we consider the situation where the target distribution is near-Gaussian. 
The rationale here is that we can first derive the optimal algorithm parameters  including $T$ by assuming the target distribution is exactly Gaussian, 
and it should be sensible to use  derived parameter values for the actual target distribution
provided it does not deviate too far from Gaussian. 
Such near-Gaussian distributions arise frequently in Bayesian inference problems, especially 
for those with a large amount of data, thanks to the asymptotic normality property~\cite{gelman2013bayesian}.
Following this idea we shall assume that the target distribution can be well approximated by a multivariate Gaussian~$\pi(\-x)\approx\mathcal{N}(\mu,\Sigma)$, where $\mu$ and $\Sigma$ are the mean and the covariance of $\pi(\-x)$ respectively.  
Without loss of generality we shall assume $\mu=0$ to keep the calculations simple. 

\textit{The main result.} In this case, we aim to determine the following algorithm parameters: the momentum covariance 
matrix $M$, and
 and the integration time $T$.
Here for the convenience of theoretical analysis, we need to impose a constraint on $M$: it commutes with $\Sigma$,
and that is, we shall choose $M$ from $\mathcal{M}^+_\Sigma$, which is the class of all positive definite matrices that commutes with $\Sigma$.
As is discussed in Section~\ref{sec:mce} these parameters will be determined by solving
\begin{equation}
\max_{\{M \in \mathcal{M}^+_\Sigma,T\in R^+\}}\mathbb{E}_{\-x_0}[\log\det\mathrm{Cov}[\-x_T|\-x_0]], \label{e:minconvar}
\end{equation} 
where $R^+$ denotes all positive real numbers. 
The following theorem states the main result of the work:
\begin{figure}[!htbp]
  \centering
  \includegraphics[width=.95 \linewidth]{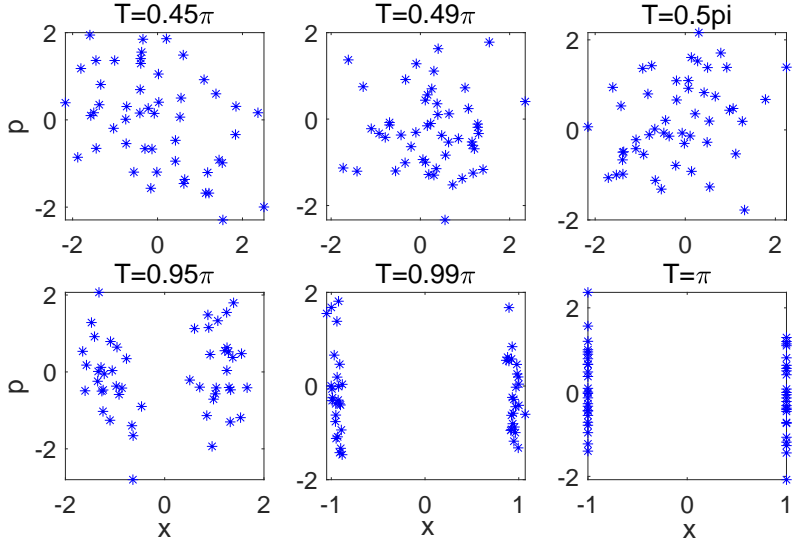}
  \caption{The scatter plots of the samples in the $(x,p)$ space. Top: the results 
  of the HMC with three different values of $T$ near the optimal value with respect to CE; Bottom: 
  the results with three different values of $T$ near the optimal value with respect to ESJD.} \label{f:eg1}
\end{figure}
\begin{theorem}
Suppose that $\pi(\-x) = \mathcal{N}(0,\Sigma)$, $\-x_0\sim \pi(\-x)$, $\-p_0\sim \mathcal{N}(0,M)$, and $\-x(t)$ is the solution of the Hamiltonian system
 \eqref{eq:hamiltonian} with $K(\-p)$ given by Eq.~\eqref{e:kp}. 
Let $\-x_T=\-x(T)$, and a solution of the optimization problem~\eqref{e:minconvar} is
\begin{equation}
M=\Sigma^{-1},\quad\mathrm{and}\quad T=(2m+1)\pi/2,
\end{equation}
for an arbitrary non-negative integer $m$. 
\end{theorem}

\textit{Proof:} First we define ${A=M^{-1}\Sigma^{-1}}$, and since $M^{-1}$ and $\Sigma^{-1}$ commutes,  $A$ is also a positive definite matrix.
Now suppose that we conduct an eigenvalue decomposition of $A$, yielding ${A}={V\Lambda V^{-1}}$ , where ${V}$ is a square matrix whose the $i$-{th} column is the eigenvector $\pmb{v}_i$ of $\pmb{A}$ and $\pmb{\Lambda}$ is the diagonal matrix whose diagonal elements are the corresponding eigenvalues, $\pmb{\Lambda}_{i,i}=\lambda_{i}$. 

Under the assumption that the target distribution $\pi(\-x)= \mathcal{N}(0,\Sigma)$, the Hamiltonian system (3) can be solved analytically, 
\begin{subequations} \label{e:gensol}
\begin{align}
\pmb{x}(t)&=\sum_{i=1}^n (a_i\cos \sqrt{\lambda_i }t+b_i\sin\sqrt{\lambda_i }t)\pmb{v}_i, \label{e:xt}\\
\pmb{p}(t)&={M}\sum_{i=1}^n \sqrt{\lambda_i} (-a_i\sin \sqrt{\lambda_i }t+b_i\cos\sqrt{\lambda_i }t)\pmb{v}_i,
\end{align}
where the coefficients $\{a_i,b_i\}_{i=1}^n$ are determined via the initial conditions:  
\begin{equation}
\label{eq:ini}
\pmb{x}_0=\sum_{i=1}^n a_i\pmb{v}_i,\quad
\pmb{p}_0={M}\sum_{i=1}^n \sqrt{\lambda_i}b_i\pmb{v}_i.
\end{equation}
\end{subequations}
By some elementary calculations,  we obtain from Eqs.~\eqref{e:gensol} that, 
\begin{equation}
C=\mathrm{Cov}(\pmb{x}_T|\pmb{x}_0)
=\sum_{i=1}^n\sum_{j=1}^n \mathbb{E}[b_ib_j]\sin (\sqrt{\lambda_i }T)\sin (\sqrt{\lambda_j }T)\pmb{v}_i\pmb{v}_j^T. \label{e:cov}
\end{equation}
Now using Eq.~\eqref{eq:ini} and the facts that $\pmb{x}_0\sim \mathcal{N}(0,\Sigma)$ and $\-p_0\sim\mathcal{N}(0,M)$, 
we derive that $(b_1,...,b_n)$ follows a multivariate Gaussian distribution and 
\begin{equation}
\mathbb{E}[b_i,b_j]=\Gamma_{i,j},\quad\mathrm{where}\quad \Gamma= {\Lambda^{-\frac{1}{2}}V^{-1}M^{-1}V^{-T}\Lambda^{-\frac{1}{2}}},
\label{e:bibj}\end{equation}
 for any $0\leq i,j\leq n$.

Substituting Eq.~\eqref{e:bibj} into Eq.~\eqref{e:cov} yields, 
\begin{equation}
C_{i,j}=\sum_{i'=1}^n\sum_{j'=1}^n {\Gamma}_{i',j'}\sin(\sqrt{\lambda_{i'}}T)V_{i,{i'}}\sin(\sqrt{\lambda_{j'}}T){V}_{{j'},j}.
\label{e:varnd}
\end{equation}
Alternatively, we can rewrite Eq.~\eqref{e:varnd} in a matrix form:
\begin{equation}
C =V\Lambda_{\sin}\Sigma \Lambda_{\sin}V^T,\quad\\
\end{equation}
with
\begin{equation}
\Lambda_{\sin}=\mathrm{diag}[\sin\sqrt{\lambda_1}T,...,\sin\sqrt{\lambda_n}T].
\end{equation}
It follows that
\begin{equation}
\det C =\det(V\Lambda_{\sin}\Sigma \Lambda_{\sin}V^T)
=\prod^n_{i=1}\lambda_{{\Sigma},i}\frac{1-\cos 2\sqrt{\lambda_i}T}{2},
\end{equation}
where $\lambda_{\Sigma,1},...,\lambda_{\Sigma,n}$ are the eigenvalues of $\Sigma$. 
Now recall that we want to find a solution of
\begin{equation}
\begin{aligned}
&\max_{\{M \in \mathcal{M}^+_\Sigma,T\in R^+\}}\mathbb{E}_{\-x_0}[\log\det\mathrm{Cov}[\-x_T|\-x_0]]\\
= &\log\sum^n_{i=1}\frac{1-\cos 2\sqrt{\lambda_i}T}{2}
+\log\det\Sigma. 
\end{aligned}
\end{equation} 
It is obvious that the minimum of the problem is attained at
\[
\sqrt{\lambda_i} T =\frac{(2m+1)}2\pi, \quad\mathrm{for}\quad i=1...n,
\] where $m$ is an arbitrary integer. 
A special choice is that 
\[T =\frac{(2m+1)}2\pi,\]
and $\lambda_i=1$ for $i=1...n$.
It follows immediately  that matrix $A$ is identity, which implies that  
 $M=\Sigma^{-1}$. $\Box$

 
In practice, since the Hamiltonian system needs to be solved numerically,
it is certainly desirable to use smaller integration time $T$, and thus in the HMC algorithm we set $m=1$
and $T=\pi/2$. We reinstate here  that, while the trick to improve the efficiency of HMC 
by choosing $M=\Sigma^{-1}$ has long been known from an intuitive perspective~\cite{neal2011mcmc},
we are able to provide a justification for it based on  the maximum CE (MCE) principle.

\begin{algorithm}[p!]
\caption{Maximum Conditional Entropy Sampler}\label{alg:mces}

\begin{algorithmic}[1]
\Require{$U(\mathbf{x})$, $Acc_{min}$,  $N_0$, $L_0$, $L_{max}$, $N_{max}$, $N_M$, $N_L$, $\rho$, $I_{max}$}
\State Initialization: draw $N_0$ samples $\{\mathbf{x}_{1},\mathbf{x}_{2},...,\mathbf{x}_{N_0}\}$ using standard HMC sampler.
\State Estimate the sample covariance matrix $\hat{\Sigma}$ of $\{\mathbf{x}_{1},\mathbf{x}_{2},...,\mathbf{x}_{N_0}\}$;
\State Let $M=\hat{\Sigma}^{-1}$; 
\State Let $T=\pi/2$;
\State Let $Acc_{old}=0$, $L_{old}=L_0$, $L=L_0$, and $I_L=1$,$I_M=0$;
\For{$t=N_0$ to $N_{max}$}
   \State $\epsilon = T/L$;
   \State Draw $\mathbf{p}_t\sim N(0,\ M)$;
   \State $(\mathbf{x}^*,\,\mathbf{p}^*)=\mathrm{leapfrog}(\mathbf{x}_t, \mathbf{p}_t, M, U(\mathbf{x}),\epsilon, L)$; \label{st:lf}
   \State Draw $u\sim U(0, 1)$;
   \If{$u<\min \{1,\exp(H(\mathbf{x}_t,\mathbf{p}_t)-H(\mathbf{x}^*,\mathbf{p}^*))\}$}
      \State $\mathbf{x}_{t+1}=\mathbf{x}^*$;
   \Else
      \State $\mathbf{x}_{t+1}=\mathbf{x}_t$;
   \EndIf
   
   \If { $t\mod N_{L}=0$}
      \State Let $Acc$ be the average acceptance probability of the last $N_L$ samples;
      \If{$t<N_\mathrm{M}$ and ($I_M=1$ or $Acc>0$)}\label{st:maxupdate}
        \State Update the sample covariance matrix $\hat{\Sigma}$ with the last $N_L$ samples; \label{st:updt2}
        \State Let $M=\hat{\Sigma}^{-1}$; \label{st:updtM}
        \State $I_M=1$;
      \EndIf
      \If{${I_L=1}$}

        \If {$L=L_{max}$}
           \State $I_L=0$;\label{st:ifupdate1}
           \If{$Acc/L<Acc_{old}/L_{old}$ }
              \State $L=L_{old}$;
           \EndIf
        \Else
           \If{${Acc}>{Acc}_{min}$}
            \If{$Acc/L<Acc_{old}/L_{old}$ }
                 \State $I_{count}=I_{count}+1$;
                 \If{$I_{count}\geq I_{max}$}
                    \State $I_L=0$, $L=L_{old}$;\label{st:ifupdate2}
                 \EndIf
              \Else
                 \State $Acc_{old}=Acc$, $L_{old}=L$;
                 \State $I_{count}=0$;
                 \State $L=\min\{[\rho L_{old}],L_{max}\}$ \label{st:updateL}
              \EndIf
           \Else
              \State $Acc_{old}=Acc$, $L_{old}=L$;
              \State $I_{count}=0$;
              \State $L=\min\{[\rho L_{old}],L_{max}\}$.\label{st:updateL2}
           \EndIf
        \EndIf
      \EndIf
    \EndIf
\EndFor
\end{algorithmic}
\end{algorithm}
\textit{Comparison with ESJD: a univariate example.} Here we use a simple univariate example to to demonstrate the difference between the CE criterion and the ESJD.
 Namely we assume that the target distribution is  $N(0,k)$,
$K(p)=p^2/(2m)$ and $p_0\sim N(0,m)$.  In this case, by some elementary calculations we can derive that an integration time that maximizes ESJD~\eqref{e:esjd}
 is $T=\sqrt{km} \pi$, and
the associated proposal is: 
$x_T = -x_0$, which means that the samples will only jump between two locations ($x_0$ and $-x_0$) and the resulting chain is certainly not ergodic. 
On the other hand, the optimal integration time with respect to CE is $T = \sqrt{km}\pi/2$, and the associated  
proposal is $\pi(x_T|x_0)=N(0,k)$, i.e., to sample directly from the target distribution, which 
is the very optimal distribution in this case. 
We refer to the the SM for a detailed derivation of the optimal integration time with respect to KSE and ESJD.  
It is also easy to derive that, 
the efficiency of the proposal behaves periodically with respect to $T$ in this case.
In example 1, we use numerical experiments to validate the theoretical analysis conducted here.

\textit{Determining the number of leapfrog steps.}
Now we have derived the optimal integration time $T$, and as is discussed earlier, we need to integrate  
the Hamiltonian system~\eqref{eq:hamiltonian} from $0$ to $T$ using the leapfrog algorithm. 
In particular, we usually need to perform leapfrog integration multiple times to achieve the necessary numerical precision,
and to this end, the number of leapfrog steps, conventionally denoted by $L$, is another key algorithm parameter to be specified. 
If we increase $L$, the numerical integration becomes more accurate and the acceptance probability
approaches to 1, and the price to pay is that more leapfrog steps means higher computational cost. 
 We note here that 
in most existing works both $L$ and $\epsilon=T/L$ need to be determined simultaneously, and in that case higher acceptance probability does not necessarily imply a better proposal even without considering the computational cost
for computing the proposal. 
In our method, however, since the total integration time $T$ is fixed, it is reasonable to assume 
that increasing $L$, which in turn increases the acceptance probability, improves the performance of the algorithm.   
Based on this idea, we shall seek the value of $L$ that provides the highest acceptance rate per computational 
cost (which is usually measured by $L$).  
Namely, we use an adaptive scheme to gradually increase $L$ until the average acceptance rate per $L$ does not improve.

\textit{Adaptively estimating the covariance matrix.} Another important issue in the method is that it requests 
the knowledge of the target covariance matrix. 
Here we follow the idea of the adaptive MCMC algorithms to estimate the covariance from the sample history
and specifically the target covariance is updated using the method given in~\cite{haario2001adaptive,haario2006dram}. 
We emphasize here that, the adaptive algorithm does not require an accurate 
estimation of the target covariance in advance (i.e. from a pilot runs); rather it adaptively
improves the estimate of the covariance as the sample size increases~\cite{andrieu2008tutorial}.
We present the complete algorithm in Alg.~\ref{alg:mces},
and provide some remarks on the algorithm in the following:
\begin{itemize}
\item In step~\ref{st:lf}, the function $\mathrm{leapfrog}(\mathbf{x}_t, \mathbf{p}_t, M, U(x),\epsilon, L)$ represents to solve 
the Hamiltonian system~\eqref{eq:hamiltonian} specified by $M$ and $U(\-x)$
from the initial condition $\-x_t$ and $\-p_t$, using the leapfrog algorithm with stepsize $\epsilon$ for $L$ steps.
\item In step~\ref{st:maxupdate}, we fix the maximum number of iterations in which we update the parameters to be $N_M$. 
\item Steps~\ref{st:updt2} and \ref{st:updtM} update matrix $M$ from samples,  and the formulas 
for these computations are  Eqs. (2) and (3) ~\cite{haario2001adaptive} respectively;
\item In steps~\ref{st:updateL} and \ref{st:updateL2}, we update $L$ by multiplying the current value of it by a factor $\rho$.
\item In steps~\ref{st:ifupdate1} and \ref{st:ifupdate2}, we set the condition for  stopping the update of $L$:  the $Acc/L$ quantity 
does not improve in $I_{max}$ consecutive steps and the minimal acceptance probability 
$Acc_{min}$ has been reached. 

\end{itemize} 
\section{Numerical examples}\label{sec:examples}
In this section, we provide four examples to demonstrate the performance of the proposed MCE sampler (MCES). 
The purpose of the first example to demonstrate that the failure of the ESJD criterion for the simple Gaussian distribution while  
MCE performs well;
the second example is used to test how the method performs at several different levels of 
non-Gaussianity; 
finally the last two examples  provide real-world testbeds in which we compare 
the performance of the MCE method and NUTS. 
The code of the proposed MCE method and the examples provided here are available at 
the Github repository~\cite{mcesmatlab}. 

\subsection{Toy Problem: Univariate Gaussian}
We first use a Gaussian example  
 to compare the CE criterion with ESJD.
Specifically we take target distribution to be the univariate standard Gaussian distribution $N(0,1)$,
the kinetic energy to be $K(p)=p^2/2$, and $p_0$ to be standard Gaussian as well. 
As has been discussed, the optimal integration time $T$ computed by the proposed MCE method is $\pi/2$,
 while that computed by ESJD is $\pi$. 
We thus take six different values of $T$: $T=0.45\pi,\,0.49\pi,\,0.5\pi$, which are close to the optimal value by MCE,
and $T=0.95\pi,0.99\pi,\,\pi$, which are close to the optimal value predicted by ESJD.
In all the tests we fix the starting position to be $x=1$, and since for this toy problem, we have the analytical solution (see Appendix A) and so we do not need to use leapfrog. 
To demonstrate the behavior of the proposals, we plot the first 100 samples produced by each proposal in Fig.~\ref{f:eg1}.
First it can be seen here that if we use $T=\pi$ which is exactly the optimal value predicted by ESJD, 
the samples are fixed at $x=-1$ and $x=1$ and in this case the sampler fails completely. The samples deviate away from the two points as $T$ is slightly perturbed from $\pi/2$, but most samples are still concentrated near the two locations, 
indicating poor performance of the proposals. 
On the other hand, the samples drawn by the proposals determined with the MCE method 
distribute well according to the target distribution, which demonstrates that
the integration time predicted by the MCE method does lead to good proposals in this Gaussian example.

\begin{table}[!htbp]
 \centering
 \small
 \begin{tabular}{|c|c|c|c|c|c|}
  \hline
  Parameter&$L_0$&$L_{max}$ &$\rho$ &$Acc_{min}$ &$N_{max}$\\
  \hline
  Value &1 &60&1.2&$60\%$&10000\\
  \hline
  Parameter&$N_{L}$ &$N_{M}$ &$N_{0}$& $I_{max}$&\\
  \hline
  Value&200&2000&1000&1&\\
        \hline
 \end{tabular}
 \medskip
 
 \caption{Algorithm parameters of MCES.}\label{tb:params}
\end{table}

\subsection{Toy Problem: Rosenbrock function}
\begin{figure}
  \centering
	\includegraphics[width=.95\linewidth]{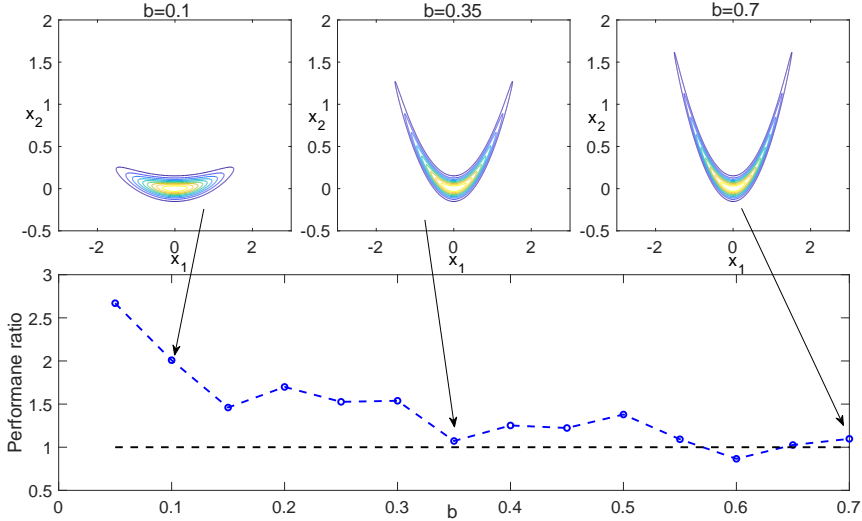}
	\caption{Top: the Rosenbrock distributions with $b=0.1,\,0.35$, and $0.7$ from left to  right. Bottom: 
	The combined $ESS/L$ ratio plotted against the value of $b$.
	The black dashed line indicates a reference where the ratio is 1.} \label{f:eg2}
\end{figure}

Our second example is the Rosenbrock function, an often used benchmark problem for MCMC methods.
Specifically, we here use a slightly modified version of the Rosenbrock function:
\begin{equation}
\pi(x_1,x_2)\propto\exp(-x_1^2-100(x_2-bx_1^2)^2), 
\end{equation} 
where the modification allows us to control the ``non-Gaussianity'' of the distribution. 
Specifically the distribution is exactly Gaussian for $b=0$ and it departs away
from Gaussian if we increase the value of $b$. 
In Fig.~\ref{f:eg2} (Top), we show the distributions for $b=0,\,0.35$ and $0.1$ respectively,
where we can see that the distribution departs more from Gaussian as $b$ increases,
and it becomes significantly non-Gaussian for $b$ is larger than $0.35$.
Here we shall compare the performance of MCES and NUTS   
for various values of $b$ ranging from 0.05 to 0.7, and evaluate how the non-Gaussianity affacts the performance of MCES. 
 To do so, for each method we repeat the simulations 100 times, and in  each simulation we draw $10000$ samples with additional $1000$ samples use in the burn-in period (the setup is also used in Examples 2 and 3).
The algorithm parameters of MCES used in this examples and the following two are shown in Table~\ref{tb:params}. 
  In this work NUTS is implemented using the Matlab package~\cite{nutsmatlab} by Nishimura.
We then compute the Effective Sample Size (ESS) per $L$ of each simulation, which is an often used 
 performance measure of MCMC algorithms. 
The ESS is also computed using the code in \cite{nutsmatlab}:
$$ESS=\frac{n}{(1+2\sum_{k=1}^{+\infty}\rho(k))},$$
where $n$ is the number of samples and $\rho(k)$ is the auto-correlation of lag $k$.
For better graphical illustration of the performance, we define the following performance 
ratio: namely suppose that the average $ESS/L$ of MCES and NUTS are respectively $E_{1}$ 
and $E'_1$ for dimension $x_1$,  and $E_{2}$ and $E'_2$ for $x_2$, and the performance ratio
is calculated as,
\[ \mathrm{Ratio}= \frac12(E_1/E'_1+E_2/E'_2).
\]
It should be clear that the performance ratio being larger than 1 indicates that 
MCES has better performance than NUTS in terms of $ESS/L$, and it being less than one indicates the opposite. 
The performance ratio is plotted as a function of $b$ in Fig.~\ref{f:eg2} (bottom),
and from the figure we can see that the ratio 
is  considerably than one in the range around $0.05$ to $0.3$, 
and is still close to one for $b$ from $0.3$ to $0.7$.
The results suggest that the performance of MCES does decrease as the target distribution becomes more non-Gaussian;
however we also can see that 
 MCES still yields comparable performance of NUTS in that regime,
suggesting MCES may possibly be applied to problems that are considerably non-Gaussian. 
That said, we acknowledge that the performance of the method for strongly non-Gaussian distributions should be more 
throughly tested.  

\begin{figure}
  \centering
	\includegraphics[width=.48\linewidth]{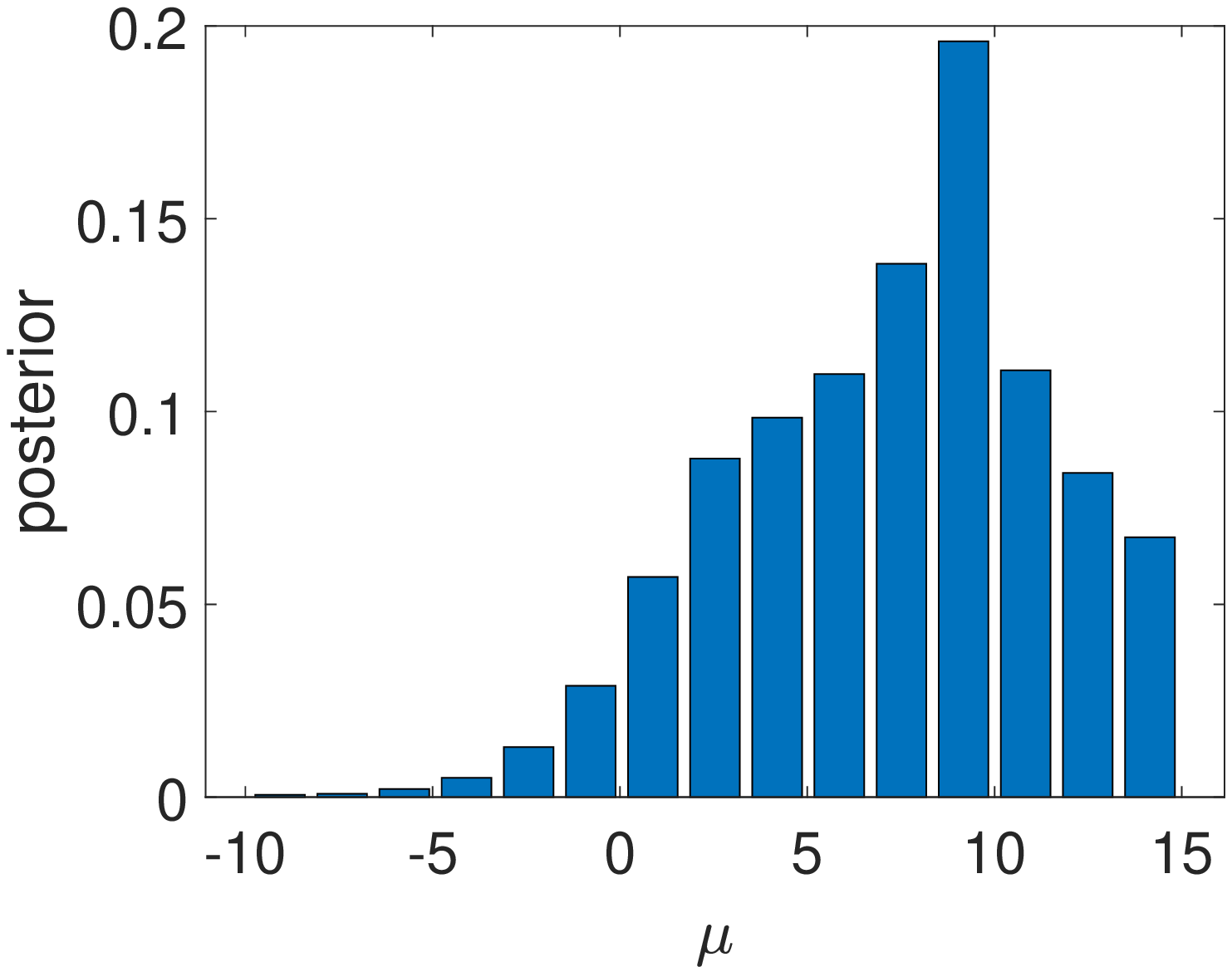}
		\includegraphics[width=.48
    \linewidth]{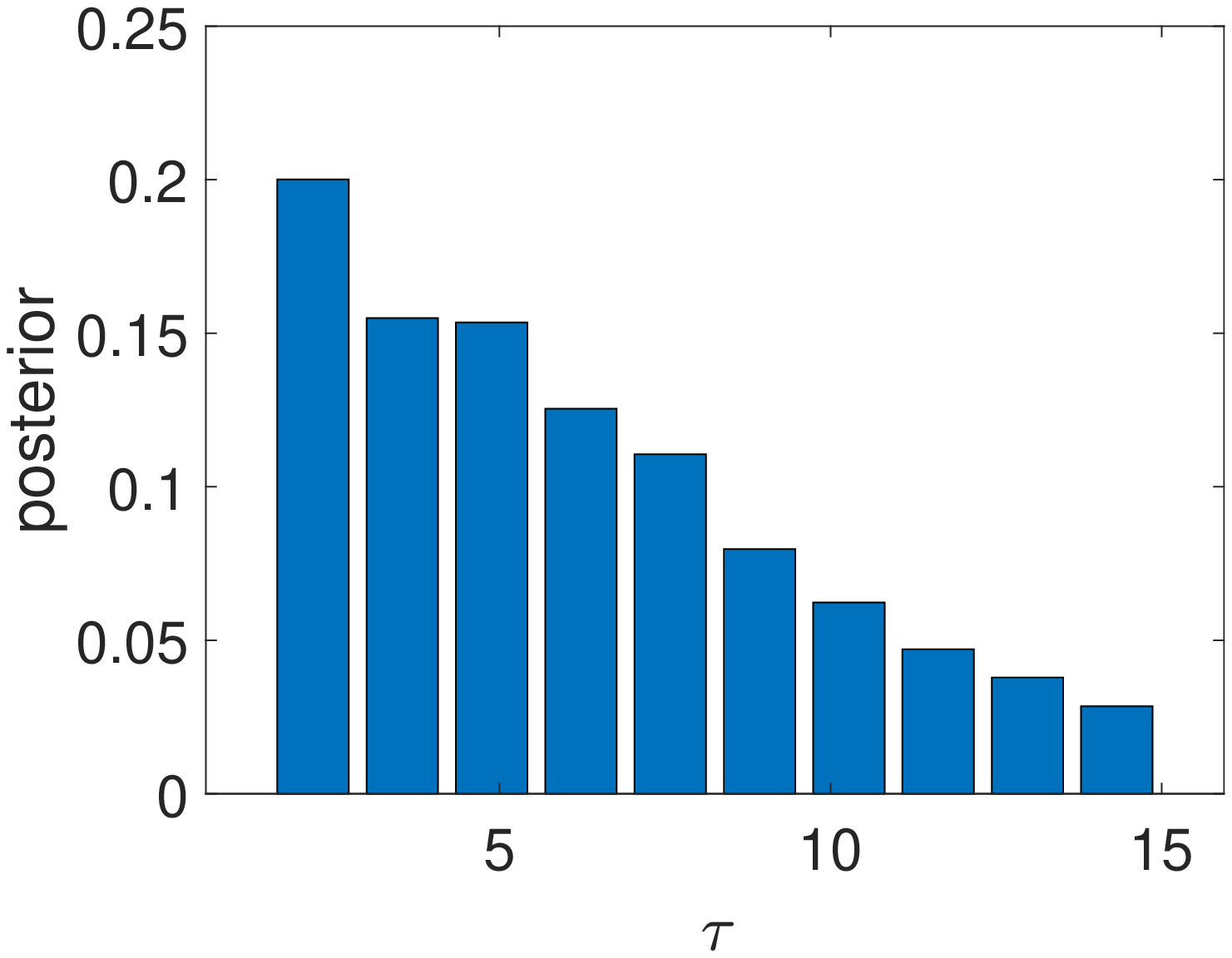}
	\caption{The histograms of the posterior samples for $\mu$ (left) and $\tau$ (right).} \label{f:eg3_pdf}
\end{figure}
\begin{table*}[htbp!]

 \centering
     \begin{tabular}{lcccccccccc}
  \hline

  \hline
  & $\theta_1$ & $\theta_2 $ & $\theta_3 $ & $\theta_4 $  & $\theta_5 $ &  $\theta_6$ & $\theta_7$ & $\theta_8$
  & $\mu$ &$\tau$   \\ \hline
  MCES & $10.3$ & $ 7.5$ & $ 6.0$ & $7.3$ & $ 5.0 $ & $ 6.0$& $ 10.0$ & $7.8$ & $ 7.3$& $5.7$\cr
   & $\left(7.1\right) $ & $ \left(5.8\right) $ & $ \left(6.9\right)$ & $ \left( 6.1\right)$ & $ \left(5.9\right) $ &$\left(6.2\right)$ & $\left(6.1\right)$ & $ \left(7.0\right) $ & $\left(4.2\right)$&  $\left(3.6\right)$\\\hline
  NUTS   &$10.1$ &$7.4$& $6.0$& $7.2$ & $5.1$ & $6.0$ & $9.8$&$7.7$ & $7.2$& $5.5$\cr
  & $\left(7.0\right) $& $\left(5.8 \right) $ &    $\left(6.8\right) $ & $\left(6.0\right) $ & $\left(5.8\right) $ & $\left(6.1\right)$& $\left(6.1\right)$& $\left(6.8\right)$& $\left(4.2\right)$ & $\left(3.7\right)$\\\hline
     \end{tabular}
   \medskip

   \caption{The posterior mean and the standard deviation (in parenthesis) computed by MCES and  NUTS for the Eight School example. } \label{tb:eg2}
\end{table*}
\subsection{Eight School problem}\label{sec:8school}
\begin{figure}
  \centering
  \includegraphics[width=.95\linewidth]{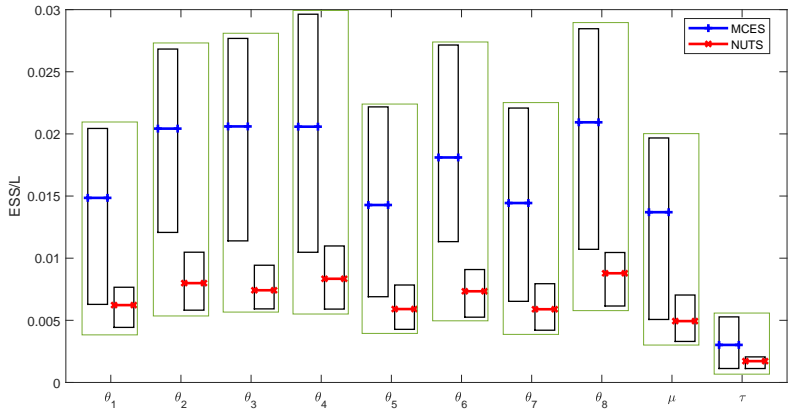}
  \caption{The box plots of the ESS per $L$ for both the MCE sampler (MCES) and the NUTS.} \label{f:eg3}
\end{figure}

Our third example is the Eight School problem in~\cite{gelman2013bayesian}, which is 
a hierarchical Bayesian inference application. While referring interested readers to \cite{gelman2013bayesian}, we here omit all the application background
and proceed directly to the mathematical setup of the problem.
Specifically, let $\{\theta_1,...,\theta_8\}$ be the parameters of interest, and $\{(y_1,\sigma_1),...,(y_8,\sigma_8)\}$ be the data. 
Let $\mu$ and $\tau$ be the hyperparameters specifying the prior of $\theta_1,\ldots,\theta_8$. The hierarchical model is:
\begin{align*}
 &\mu \sim \mbox{Uniform[-15,15]},\quad 
 \tau \sim \mbox{Uniform[0,15]}\\
 &\theta_i \sim \mathcal{N}(\mu,\tau),\quad
 y_i \sim \mathcal{N}(\theta_i,\sigma_i), \quad i=1...8.
\end{align*}
 We use the same data as those in \cite{gelman2013bayesian} in the inference problem, and we sample the posterior distribution
with the MCE sampler and the NUTS. 
First to validate the proposed method, we draw $10^5$ samples from the posterior distribution
using both the MCE sampler and NUTS, and the posterior means and variances for all the parameters obtained by both methods 
are reported in Table~\ref{tb:eg2}, from which we can see that the results computed by both methods agree well with each other,
up to certain statistical errors.
We then plot the obtained posterior histograms for $\mu$ and $\tau$ in Fig.~\ref{f:eg3_pdf},
and we can see from the figure that both distributions are significantly apart from Gaussian.
Next we compare the performance of the MCE sampler and NUTS. 
We compute the ESS$/L$ for the results of each simulation, which is used as a performance measure of the samplers,
and we show the box-plots of the ESS$/L$ results in Fig.~\ref{f:eg3}. 
The plots show that the MCE sampler achieves evidently higher ESS per $L$ than NUTS,
even for the two dimensions that are evidently non-Gaussian. 
Additional test results (using different algorithm parameters) are provided in Appendix~\ref{sec:moreresults},
and the results demonstrate similar performance to those presented in this section, 
suggesting that MCES is rather robust against different values of algorithm parameters. 

\subsection{Bayesian Logistic regression with the German credit data}
Our last example is the German credit data available at~\cite{Dua:2019}, a popular benchmark problem for Logistic regression. 
Simply put, this problem aims to classify people described by a set of attributes as good or bad credit risks.
Here we use the modified version with all numerical attributes~\cite{Dua:2019}, 
which has 1000 instances each with a 24-dimensional numerical input and a binary output. 
For further details of the dataset, please refer to the description of the data set at \cite{Dua:2019}. 
Here the problem is formulated as a Logistic regression~\cite{hosmer2013applied} and the regression coefficients 
$\bm{\beta} = (\beta_0,\beta_1,...,\beta_n)$ with $n=24$ are estimated 
with a Bayesian inference. 
The prior distribution is chosen to be standard Gaussian: $\pmb{\beta}\sim\mathcal{N}(0,{I})$. 
Just like the previous example, for each method we repeat the simulations 100 times, and in each simulation we draw $10000$ samples with additional $1000$ samples used for burn-in.
The algorithm parameters are the same as those used in Example 2.  
To validate the MCES method, we have verified that the posterior means and variances computed 
by both methods agree well with each other.
Specifically in Table~\ref{tb:ega4} we present the posterior means and variances of all the parameters computed 
by MCES and NUTS. Just like the Eight School example, the results of the two methods agree well with each other, up to certain 
statistical errors. 

We estimate the ESS$/L$ of each dimension for  each simulation, and we then compute the mean and the standard deviation of the 100 trials.
We show the results in Table~\ref{tb:eg4}, 
which show that the average $ESS/L$ of MCES is higher than twice of that of NUTS 
in all the 25 dimensions.
These results demonstrate that the MCES method has a good performance in the
Bayesian Logistic regression, a class of often-encountered real world problems. 
The good performance of the MCES method in such problems is somehow as expected, since the posteriors in the Bayesian 
logistic regression usually do not deviate vastly from Gaussian~\cite{jaakkola1997variational}.

\begin{table}[htbp!]

 \centering
     \begin{tabular}{lcccccccc}
  \hline

  \hline
  {parameters} & $\beta_0$ & $\beta_1 $ & $\beta_2 $ & $\beta_3 $  & $\beta_4$ & $\beta_5$ & $\beta_6$ \\ \hline

  MCES & $-1.20$ & $-0.73$ & $ 0.42$ & $-0.41$ & $ 0.13 $ & $ -0.36$& $ -0.17$\cr

   & $\left(0.09\right) $ & $ \left(0.09\right) $ & $ \left(0.10\right)$ & $ \left( 0.09\right)$ & $ \left(0.10\right) $&$\left(0.09\right)$ & $\left(0.09\right)$ \\\hline

  NUTS   & $-1.20$ & $-0.73$ & $ 0.42$ & $-0.41$ & $ 0.13 $ & $ -0.36$& $ -0.17$\cr

  & $\left(0.09\right) $& $\left(0.09 \right) $ &    $\left(0.10\right) $ & $\left(0.10\right) $ & $\left(0.11\right) $ & $\left(0.09\right)$& $\left(0.09\right)$\\\hline

  {parameters} & $\beta_7$ & $\beta_8$ & $\beta_9$  &$\beta_{10}$ & $\beta_{11}$ & $\beta_{12} $& \\ \hline

  MCES  & $-0.15$ & $ 0.01$& $0.18$& $-0.11$ & $ -0.22$ & $ 0.12$&\cr

  & $ \left(0.08\right) $ & $\left(0.09\right)$&  $\left(0.10\right)$& $\left(0.10\right) $ & $ \left(0.08\right) $ & $ \left(0.09\right)$ &\\\hline

  NUTS    & $-0.15$ & $ 0.01$& $0.18$& $-0.11$ & $ -0.22$ & $ 0.12$ &\cr

  & $\left(0.08\right)$& $\left(0.09\right)$ & $\left(0.09\right)$& $\left(0.10\right) $& $\left(0.08 \right) $ &    $\left(0.09\right) $ &\\\hline

  {parameters} & $\beta_{13} $ & $\beta_{14} $&$\beta_{15} $ &  $\beta_{16}$ & $\beta_{17}$ & $\beta_{18}$&\\ \hline

  MCES  & $0.03$ & $ -0.13 $ & $ -0.29$& $ 0.28$ & $-0.30$ & $ 0.30$&\cr

   & $ \left( 0.09\right)$ & $ \left(0.09\right) $ &$\left(0.12\right)$ & $\left(0.08\right)$ & $ \left(0.10\right) $& $\left(0.12\right)$&\\\hline

  NUTS   & $0.03$ & $ -0.14 $  & $ -0.29$& $ 0.28$ & $-0.30$ & $ 0.30$&\cr

  & $\left(0.09\right) $ & $\left(0.09\right) $& $\left(0.12\right)$& $\left(0.08\right)$& $\left(0.10\right)$& $\left(0.12\right)$ &\\ \hline

  {parameters}  &$\beta_{20}$& $\beta_{21}$ & $\beta_{22} $ & $\beta_{23} $ & $\beta_{24} $& $\beta_{19}$&\\ \hline

  MCES  &$0.12$ &$-0.06$& $-0.09$& $-0.03$ & $-0.02$ & $0.27$ & \cr

    & $\left(0.14\right) $ & $ \left(0.14\right) $ & $ \left(0.09\right)$ & $ \left( 0.13\right)$ & $ \left(0.12\right) $ & $\left(0.11\right)$&\\\hline

  NUTS   &$0.12$ &$-0.06$& $-0.09$& $-0.03$ & $-0.02$ & $0.27$&\cr

   & $\left(0.14\right) $& $\left(0.14 \right) $ &    $\left(0.09\right) $ & $\left(0.13\right) $ & $\left(0.12\right) $&$\left(0.11\right)$&\\\hline
     \end{tabular}
   \medskip

   \caption{The posterior mean and the standard deviation (in parenthesis) computed by MCES and  NUTS for 
   the German Credit example. } \label{tb:ega4}
\end{table}

\begin{table}[htbp!]
 \centering

     \begin{tabular}{lccccccccccc}
  \hline

  \hline
  {parameters} & $\beta_0 $& $\beta_1$ & $\beta_2 $ & $\beta_3 $ & $\beta_4 $\\ \hline

  MCES & $0.1314$ & $0.1337$ & $ 0.1399$ & $0.1315$ & $ 0.1426 $    \cr
  
   & $\left(0.0528\right) $ & $ \left(0.0622\right) $ & $ \left(0.0507\right)$ & $ \left( 0.0569\right)$ & $ \left(0.0540\right) $\\\hline
  
  NUTS   & $0.0554$ & $0.0532$ & $ 0.0524$ & $0.0520$ & $ 0.0519 $  \cr
  
  & $\left(0.0049\right) $& $\left(0.0036\right) $ &    $\left(0.0033\right) $ & $\left(0.0037\right) $ & $\left(0.0035\right) $ \\\hline

  {parameters} & $\beta_5 $ &  $\beta_6$ & $\beta_7$& $\beta_8$& $\beta_{9} $   \\ \hline

  MCES& $ 0.1344$& $ 0.1406$& $0.1405$ & $0.1329$ & $ 0.1373$  \cr
  
   &$\left(0.0505\right)$ & $\left(0.0539\right)$& $\left(0.0479\right) $ & $ \left(0.0553\right) $& $ \left(0.0529\right)$\\\hline

  NUTS& $ 0.0530$& $ 0.0524$ & $0.518$  & $ 0.0523$ & $0.0521$\cr

  & $\left(0.0033\right)$& $\left(0.0032\right)$& $\left(0.0034\right) $& $\left(0.0034 \right) $&    $\left(0.0034\right) $ \\\hline

  {parameters} & $\beta_{10} $ & $\beta_{11} $  & $\beta_{12} $ &  $\beta_{13}$ & $\beta_{14}$\\ \hline

  MCES  & $0.1272$ & $ 0.1327 $ & $ 0.1302$& $ 0.1414$& $0.1341$ \cr

     & $ \left( 0.0533\right)$ & $ \left(0.0505\right) $ &$\left(0.0580\right)$ & $\left(0.0546\right)$& $\left(0.0522\right) $  \\\hline

  NUTS  & $ 0.0526$ & $0.0524$ & $ 0.0519 $   & $0.0522$ & $ 0.0522 $ \cr

   & $\left(0.0033\right) $ & $\left(0.0032\right) $ & $\left(0.0034\right)$& $\left(0.0029\right)$& $\left(0.0033\right) $\\\hline

  {parameters} & $\beta_{15}$ & $\beta_{16} $ & $\beta_{17} $& $\beta_{18} $& $\beta_{19} $     \\ \hline

  MCES & $0.1294$ & $ 0.1327$ & $0.1331$  & $ 0.1346 $ & $0.1387$\cr

    & $ \left(0.0517\right) $ & $ \left(0.0544\right)$ & $ \left( 0.0534\right)$ & $ \left(0.0504\right) $ &$\left(0.0539\right)$\\\hline

  NUTS& $ 0.0543$& $ 0.0525$&$0.0532$  & $ 0.0521$& $ 0.0517$\cr

   & $\left(0.0033 \right) $ &    $\left(0.0033\right) $ & $\left(0.0034\right) $& $\left(0.0034\right) $ & $\left(0.0033\right)$\\\hline

  {parameters} &  $\beta_{20}$ & $\beta_{21}$& $\beta_{22}$ & $\beta_{23} $ & $\beta_{24} $   &  &   \\ \hline

  MCES & $ 0.1418$& $0.1392$ & $ 0.1312 $ & $ 0.1449$& $ 0.1402$&& \cr

    & $\left(0.0543\right)$& $ \left( 0.0563\right)$ & $ \left(0.0576\right) $ &$\left(0.0514\right)$ & $\left(0.0526\right)$&& \\\hline

  NUTS  &$0.0518$& $0.0522$ & $ 0.0525 $ & $ 0.0522$& $ 0.0523$&& \cr

   & $\left(0.0037\right)$& $\left(0.0035\right) $ & $\left(0.0031\right) $ & $\left(0.0037\right)$& $\left(0.0039\right)$&&\\\hline

   \end{tabular}
   \medskip

   \caption{The means and standard deviations of $ESS/L$ for MCES and  NUTS in German Credit Example.
	} \label{tb:eg4}
\end{table}

\subsection{Log-Gaussian Cox Process}
The Log-Gaussian Cox process (LGCP) is a widely used  model for spatial point process data~\cite{moller1998log}.
Mathematically it is a hierarchical structure consisting of a Poisson point process with a random log-intensity given by a Gaussian random field.
In practice, the Bayesian method is often used to infer the intensity function from the observation data,
and the resulting Bayesian inference problems have application in fields such as ecology, geology, seismology, and neuroimaging~\cite{teng2017bayesian}.
Sampling the posterior distribution for the LGCP model is computationally challenging largely due to the high dimensionality of such problems. 
In this example we consider a two-dimensional spatial Bayesian inference problem with the LGCP model. 
Let $\Omega$ be a two dimensional $d\times d$ spatial grid, indexed by $\{(i,j)|i,j=1,...,d\}$,
and let $\pmb{X}=\{x_{i,j}\}_{i,j=1}^d$ be a Gaussian process defined on $\Omega$:
namely the elements in any nonempty subset of $\pmb{X}$ follows a (multivariate) Gaussian distribution. 
Moreover, we shall assume that the Gaussian process $\pmb{X}$ is of a constant mean $\mu$ and a squared exponential 
covariance function:
 $$\Sigma[(i,j),(i',j')]=\alpha\exp(-\delta(i,i',j,j')/\beta d), \,\,\mathrm{with}\,\, \delta(i,i',j,j')=\sqrt{(i-i')^2+(j-j')^2},$$
 where $\alpha$ and $\beta$ are parameters that will be specified later. 
 Next  from $\pmb{X}$, we define the latent intensity process $\Lambda=\{\lambda_{i,j}\}$ with means $\lambda_{ij}=s\exp(x_{ij})$ where 
 $s$ is a scalar parameter.
Let  $\pmb{Y}=\{y_{ij}|i,j,=1,...,d\}$ be the data set that are observed where data points $y_{i,j}$ are Poisson distributed with mean $\lambda_{i,j}$:
\[y_{i,j}\sim \mathrm{Poisson}(\cdot,\lambda_{i,j}),\]
 and conditionally independent given the latent intensity process $\Lambda=\{\lambda_{i,j}\}$. 
 The goal of the problem is to compute the posterior distribution of $\pmb{X}$ given the data set $\pmb{Y}$. 
 
\begin{figure}[!htbp]
  \centering
  \includegraphics[width=0.95 \linewidth]{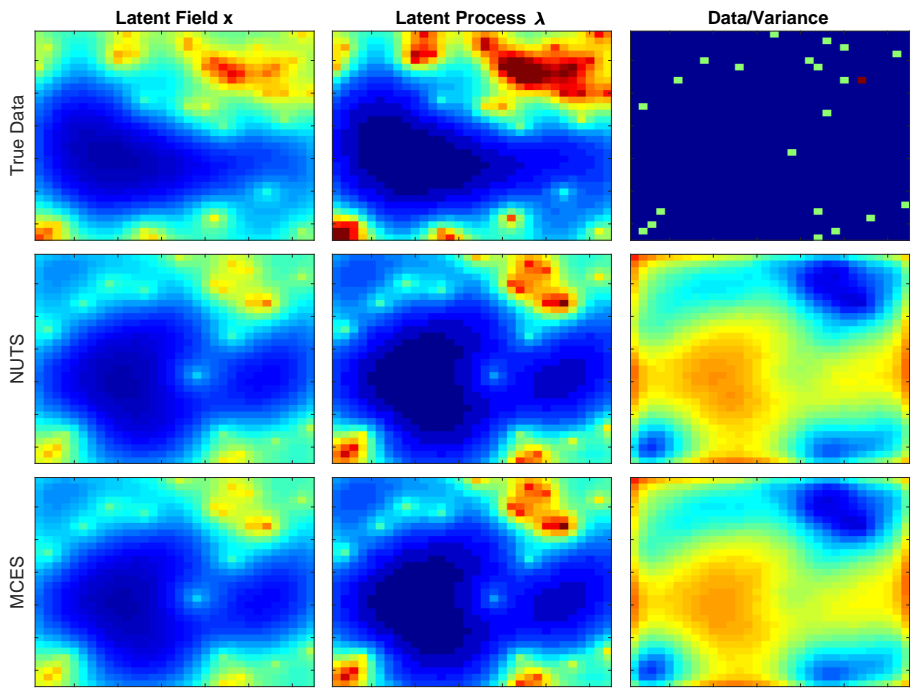}
  \caption{Comparing quality of posterior distributions from samples obtained using NUTS and MCES for the LGCP model. The top-right image shows the observation data.} \label{f:cox}
\end{figure}

\begin{figure}[!htbp]
  \centering
  \includegraphics[width=1.0 \linewidth]{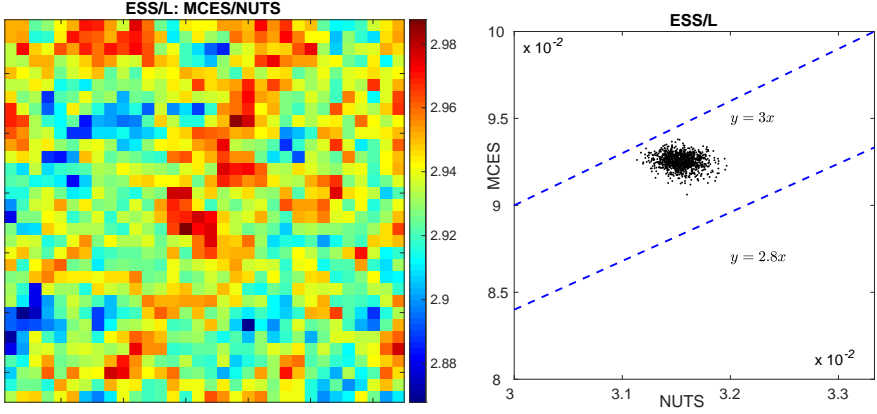}
  \caption{The $ESS/L$ comparison of MCES and NUTS for the LGCP model.
  Left: the ratio between the ESS/L value of MCES and that of NUTS at each location.
  Right: the scatter plot of the ESS/L of the two methods of all the dimensions.} \label{f:cox_ess}
\end{figure}

 \begin{table}[!htbp]
 \centering
 \begin{tabular}{|c|c|c|c|c|c|}
  \hline
  Parameter&$\alpha$&$\beta$  &$\mu$ &$d$&$s$ \\
  \hline
  Value &1.91 &1/33&$\log126-\alpha/2$&32&$1/d^2$\\
        \hline
 \end{tabular}
 \medskip
 \caption{Model parameters for the LGCP example.}\label{tb:cox}
\end{table}
In the numerical tests, we specify the model parameters as is in Table~\ref{tb:cox},
and we emphasize here that we take $d=32$, resulting in a 1024 dimensional inference problem. 
Moreover in our tests, the ground truth is randomly generated from the prior and then a synthetic data set is obtained from the generative process for this model;
both the ground truth and the generated data set are shown in Fig.~\ref{f:cox}.   
We then sample the posterior distribution using both NUTS and MCES. 
As the problem is of very high dimensions, we draw $5.5\times10^5$ samples from the posterior distribution with either method,
with the first $50000$ samples are used as the burn-in in both methods. 
We emphasize here that we use such a large number of samples in this example 
because of the high dimensionality of the problem.  
We compare the posterior results of both methods in Figs.~\ref{f:cox}:
namely the first row of the figures are from left to right respectively the ground truth of the log-intensity $x$,
that of the intensity $\lambda$ and the observation data;
the second row shows the posterior mean and variance computed by NUTS, and 
the last one shows those posterior statistics computed by the proposed MCES method. 
One can see that, both the posterior statistics computed by both methods agree quite well with each other, 
and the posterior means are reasonably close to the ground truths,
which validates the samples drawn by both methods. 
Next we shall compare the sampling efficiency of the two methods, measured by $ESS/L$,
and we show the comparison results in Figs.~\ref{f:cox_ess}. 
In the left plot of Figs.~\ref{f:cox_ess} we show 
the ratio between the $ESS/L$ value of MCES and that of NUTS at each spatial point of the graph,
and we can see that at each spatial point, the ratio is near three, which means that
the $ESS/L$ value of MCES is around 3 times as much as that of NUTS at each location. 
Moreover,  the performance ratio is rather stable across locations, ranging between 2.8 to 3. 
In the right figure of Figs.~\ref{f:cox_ess}, we show the scatter plot of the ESS/L value of all the 1024 dimensions:
 each dimension is represented as a scatter point with $ESS/L$ of NUTS being the abscissa and that of MCES being 
 the ordinate.  
First we can see from the plot that all the scattering points are bounded by the two lines $y=2.8x$ and $y=3x$,
indicating that MCES is at least 2.8 times as efficient as NUTS in terms of $ESS/L$. 
In addition the figure also reveals that the $ESS/L$ results for both methods 
do not vary significantly across dimensions: for NUTS the 1024 results are between 0.031 and 0.032 and for MCEC they are from 0.090 to 0.093.

\section{Conclusions}\label{sec:conclusions}
In this work we propose a new KSE/CE based design criterion for tuning the algorithm parameters in HMC. 
 We show that the KSE/CE criterion can address some limitation of the distance based design criteria such as ESJD. 
For near-Gaussian distribution we are able to derive the analytical solution to the resulting optimization problem.
We then develop an adaptive HMC algorithm based on the results. Numerical examples demonstrate 
that the proposed method has rather good performance even when the target distributions are considerably different from Gaussian.
Several issues and limitations of the method need to be addressed in the future. 
First, Algorithm~1 terminates the adaptation after a fixed number of iterations, 
which may potentially affect the efficiency of the algorithm, if the adaptation is terminated prematurely. 
To this end, an interesting question is that whether the chain can converge without such 
a mandatory termination. 
Moreover, the most serious restriction of the method is, of course, the near-Gaussian assumption,
which makes the method unsuitable for strongly non-Gaussian distributions, e.g., those
with multiple modes. It is thus of significant interest to apply the KSE/CE criterion to strongly
non-Gaussian distributions and develop suitable HMC algorithms for them. 
We plan to investigate these problems in the future.
\appendix

\section{Derivation of the MCE and the maximum ESJD solutions}
This section provides details of the derivation of the optimal integration time with respect to CE and with respect to 
ESJD when the target distribution is  $N(0,k)$. We also take $K(p)=p^2/(2m)$ and $p_0\sim N(0,m)$. 
In this case it is easy to derive that the solution of the Hamiltonian system is 
\begin{subequations}\label{e:solution}
\small
\begin{equation}
x(t)=A\cos(\sqrt{\frac{1}{km}}t+\phi_{0}),\quad
p(t)=-A\sqrt{\frac{m}k}\sin(\sqrt{\frac{1}{km}}t+\phi_{0})
\end{equation}
and the initial conditions are
\begin{equation}
x_0=A\cos(\phi_{0}),\quad
p_0=-A\sqrt{m/k}\sin(\phi_{0}).
\end{equation}
\end{subequations}
From Eq.~\eqref{e:solution},
we obtain
\[
\begin{aligned}
x_T&=A\cos(\sqrt{\frac{1}{km}}T+\phi_{0})\\
&=[x_0\cos(\sqrt{\frac{1}{km}}T)+p_0\sqrt{\frac{k}{m}}\sin(\sqrt{\frac{1}{km}}T)].
\end{aligned}
\]
As $p_0\sim N(0,m)$, it follows immediately that
\begin{equation}
\pi(x_T|x_0) = {N}\left(x_0\cos(\sqrt{\frac{1}{km}}T), k\sin^2(\sqrt{\frac{1}{km}}T)\right). 
\label{e:xtx0}
\end{equation}
Next we shall consider the two criteria separately. 

First we consider the CE criterion, which seeks to maximize $\mathbb{H}[x_T|x_0]$, and since $\pi(x_T|x_0)$ is 
univariate Gaussian, it is equivalent to 
\begin{align}
\max_{T>0}\mathbb{E}_{\-x_0}[\log\mathrm{Var}[\-x_T|\-x_0]]:=&\log[k\sin^2(\sqrt{\frac{1}{km}}T)]\notag\\
=&\log[1-\cos(2\sqrt{\frac{1}{km}}T)]+\log(\frac{k}{2}).
\end{align}
It is easy to see that the solution is, 
\[T=\frac{1}2\sqrt{km}(\pi+2J\pi),\]
where $J$ is an arbitrary non-negative integer. 
Certainly we should take $J=0$ and so we obtain the smallest $T$ as larger $T$ implies higher computational cost 
of the numerical integration. 
Thus the optimal solution with respect to the CE criterion is
\[T=\frac{\pi}2\sqrt{km}.\]

Next we shall derive the optimal value of $T$ with respect to ESJD. 
That is we want to solve,
\[\max_{T>0}\mathbb{E}_{x_0,p_0}[||x_T-x_0||^2].\]
Once again from Eq.~\eqref{e:solution} we obtain,
\[
\begin{aligned}
x_T-x_0
&=A\cos(\sqrt{\frac{1}{km}}T+\phi_{0})-A\cos(\phi_{0})\\
&=x_0[\cos(\sqrt{\frac{1}{km}}T)-1]+p_0\sqrt{\frac{k}{m}}\sin(\sqrt{\frac{1}{km}}T).
\end{aligned}
\]
Then we have, 
\[
\mathbb{E}_{x_0,p_0}||x_T-x_0||^2=2k[1-\cos(\sqrt{\frac{1}{km}}T)].
\]
Thus, maximizing the ESJD becomes,
\[\max_{T>0} 2k\left[1-\cos(\sqrt{\frac{1}{km}}T)\right],\]
and the solution is 
\[T=\sqrt{km}(\pi+2J\pi),\]
and for the same reason as above we take $J=0$,
which yields the optimal integration time with respect to the ESJD,
\[T=\pi\sqrt{km}.\]

\section{Additional test results for the Eight School example}\label{sec:moreresults}
\begin{figure}[!htbp]
  \centering
	\includegraphics[width=.85\linewidth]{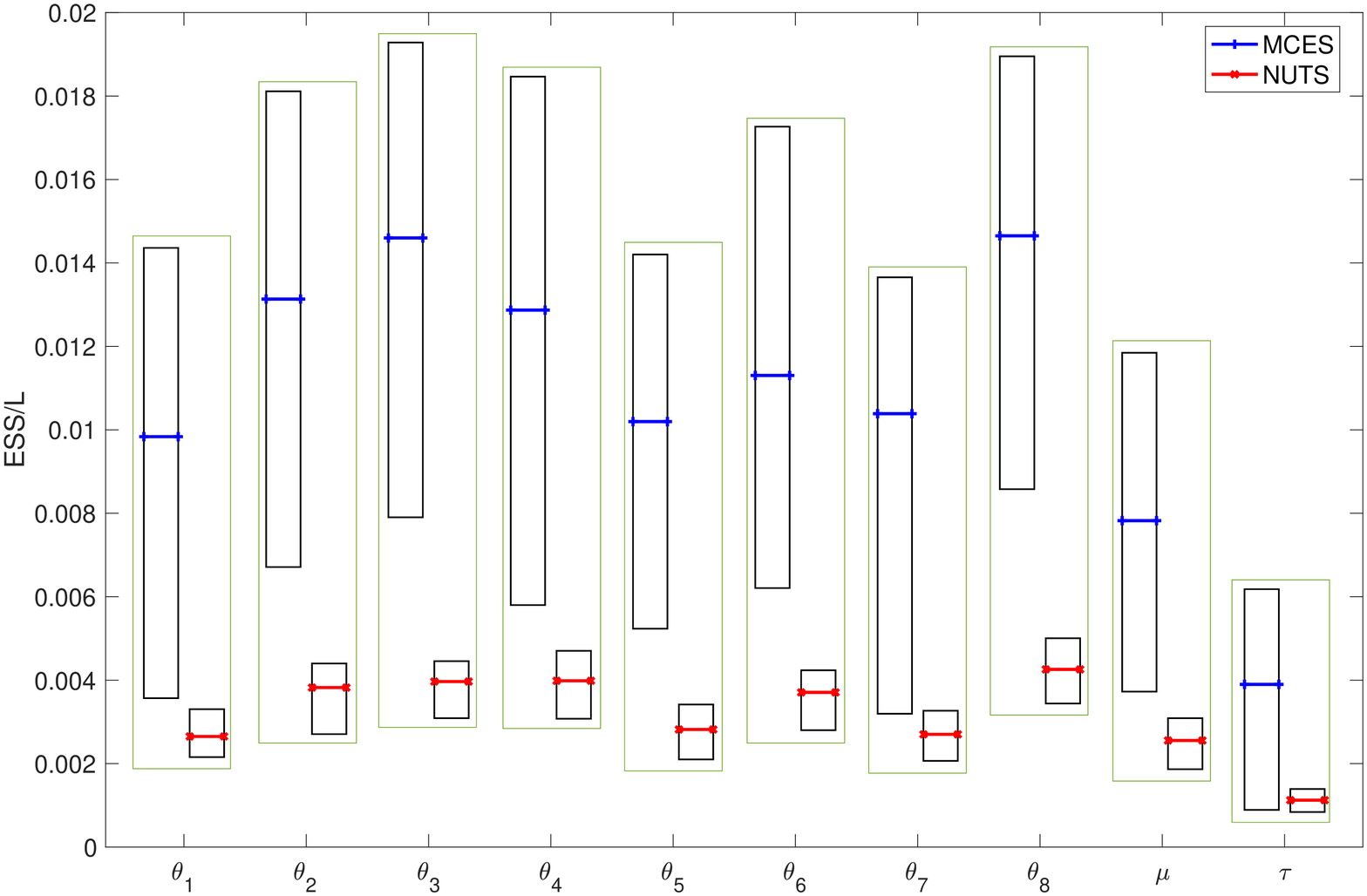}
	\caption{The box plots of the ESS per $L$ for test  1.} \label{f:ega1}
\end{figure}

\begin{figure}[!htbp]
  \centering
	\includegraphics[width=.85\linewidth]{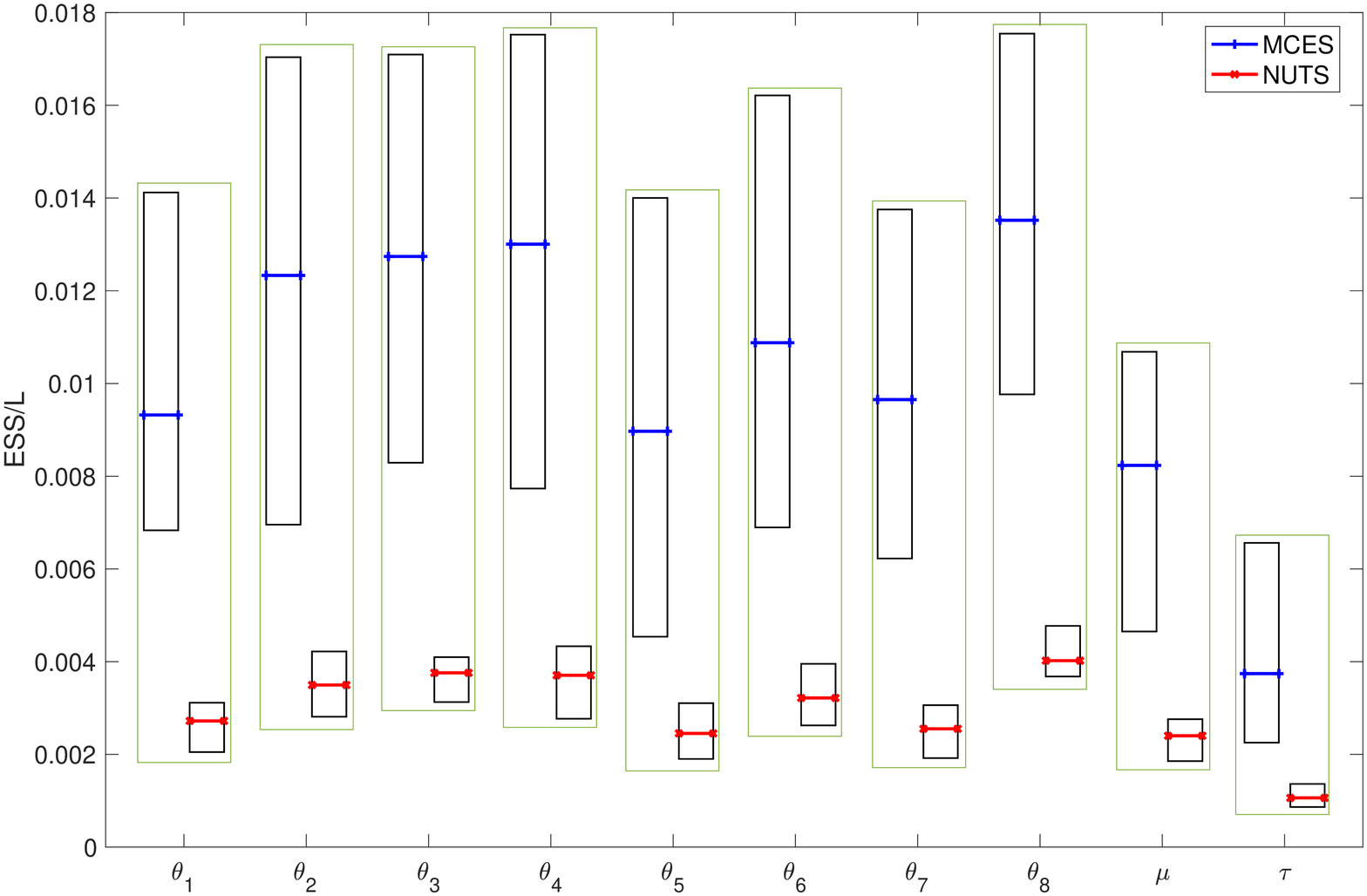}
	\caption{The box plots of the ESS per $L$ for test  2.} \label{f:ega2}
\end{figure}

\begin{figure}[!htbp]
  \centering
	\includegraphics[width=.85\linewidth]{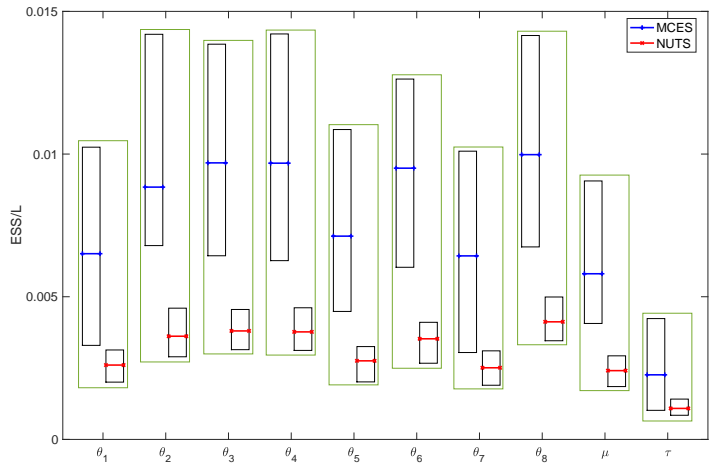}
	\caption{The box plots of the ESS per $L$ for test  3.} \label{f:ega3}
\end{figure}

We present three additional test results for the Eight School example to demonstrate 
the robustness of the method against the algorithm parameter values. 
Specifically we implemented the MCES with another three different sets of parameters shown in the tables \ref{tb:params1}-\ref{tb:params3} below. 
Compared to the parameter values used in Section~\ref{sec:8school},
we vary the values of three key parameters: 
in test 1 we change $Acc_{min}$ from $60\%$ to $40\%$,
 in test 2 we change $I_{max}$ from 2 to $1$,
and in test 3 we change $I_{max}$ from 60 to 100.
We plot the ESS$/L$ results for the three tests in 
Figs.~\ref{f:ega1}, \ref{f:ega2} and \ref{f:ega3} respectively. 
The figures demonstrate that, in all the tests,  the MCE-HMC method yields evidently better ESS per $L$ results than NUTS, suggesting that the performance of the MCE method 
is not sensitive to these parameters.

\begin{table}[!htbp]
 \centering
 \begin{tabular}{|c|c|c|c|c|c|c|c|c|c|}
  \hline
  Parameter&$L_0$&$L_{max}$ &$\rho$ &$Acc_{min}$ &$N_{max}$&$N_{L}$ &$N_{M}$ &$N_{0}$& $I_{max}$\\
  \hline
  Value &1 &60&1.2&$40\%$&10000&200&2000&1000&2\\
        \hline
 \end{tabular}
 \medskip
 \caption{Algorithm parameters for test 1.}\label{tb:params1}
\end{table}

\begin{table}[!htbp]
 \centering
 \begin{tabular}{|c|c|c|c|c|c|c|c|c|c|}
  \hline
  Parameter&$L_0$&$L_{max}$ &$\rho$ &$Acc_{min}$ &$N_{max}$&$N_{L}$ &$N_{M}$ &$N_{0}$& $I_{max}$\\
  \hline
  Value &1 &60&1.2&$60\%$&10000&200&2000&1000&1\\
        \hline
 \end{tabular}
 \medskip
 \caption{Algorithm parameters for test  2.}\label{tb:params2}
\end{table}

\begin{table}[!htbp]
 \centering
 \begin{tabular}{|c|c|c|c|c|c|c|c|c|c|}
  \hline
  Parameter&$L_0$&$L_{max}$ &$\rho$ &$Acc_{min}$ &$N_{max}$&$N_{L}$ &$N_{M}$ &$N_{0}$& $I_{max}$\\
  \hline
  Value &1 &100&1.2&$60\%$&10000&200&2000&1000&2\\
        \hline
 \end{tabular}
 \medskip
 \caption{Algorithm parameters for test 3.}\label{tb:params3}
\end{table}

\bibliographystyle{siam}
\bibliography{mces}

\begin{thebibliography}{10}

\bibitem{andrieu2008tutorial}
{\sc C.~Andrieu and J.~Thoms}, {\em A tutorial on adaptive mcmc}, Statistics
  and computing, 18 (2008), pp.~343--373.

\bibitem{barp2018geometry}
{\sc A.~Barp, F.-X. Briol, A.~D. Kennedy, and M.~Girolami}, {\em Geometry and
  dynamics for markov chain monte carlo}, Annual Review of Statistics and Its
  Application, 5 (2018), pp.~451--471.

\bibitem{beskos2013optimal}
{\sc A.~Beskos, N.~Pillai, G.~Roberts, J.-M. Sanz-Serna, A.~Stuart, et~al.},
  {\em Optimal tuning of the hybrid monte carlo algorithm}, Bernoulli, 19
  (2013), pp.~1501--1534.

\bibitem{betancourt2017conceptual}
{\sc M.~Betancourt}, {\em A conceptual introduction to hamiltonian monte
  carlo}, arXiv preprint arXiv: 1701.02434,  (2017).

\bibitem{bui2014solving}
{\sc T.~Bui-Thanh and M.~Girolami}, {\em Solving large-scale pde-constrained
  bayesian inverse problems with riemann manifold hamiltonian monte carlo},
  Inverse Problems, 30 (2014), p.~114014.

\bibitem{carpenter2017stan}
{\sc B.~Carpenter, A.~Gelman, M.~D. Hoffman, D.~Lee, B.~Goodrich,
  M.~Betancourt, M.~Brubaker, J.~Guo, P.~Li, and A.~Riddell}, {\em Stan: A
  probabilistic programming language}, Journal of statistical software, 76
  (2017).

\bibitem{chen2014stochastic}
{\sc T.~Chen, E.~Fox, and C.~Guestrin}, {\em Stochastic gradient hamiltonian
  monte carlo}, in International conference on machine learning, 2014,
  pp.~1683--1691.

\bibitem{Dua:2019}
{\sc D.~Dua and C.~Graff}, {\em {UCI} machine learning repository}, 2017.

\bibitem{duane1987hybrid}
{\sc S.~Duane, A.~D. Kennedy, B.~J. Pendleton, and D.~Roweth}, {\em Hybrid
  monte carlo}, Physics letters B, 195 (1987), pp.~216--222.

\bibitem{gelman2013bayesian}
{\sc A.~Gelman, H.~S. Stern, J.~B. Carlin, D.~B. Dunson, A.~Vehtari, and D.~B.
  Rubin}, {\em Bayesian data analysis}, Chapman and Hall/CRC, 2013.

\bibitem{girolami2011riemann}
{\sc M.~Girolami and B.~Calderhead}, {\em Riemann manifold langevin and
  hamiltonian monte carlo methods}, Journal of the Royal Statistical Society:
  Series B (Statistical Methodology), 73 (2011), pp.~123--214.

\bibitem{haario2006dram}
{\sc H.~Haario, M.~Laine, A.~Mira, and E.~Saksman}, {\em Dram: efficient
  adaptive mcmc}, Statistics and computing, 16 (2006), pp.~339--354.

\bibitem{haario2001adaptive}
{\sc H.~Haario, E.~Saksman, J.~Tamminen, et~al.}, {\em An adaptive metropolis
  algorithm}, Bernoulli, 7 (2001), pp.~223--242.

\bibitem{hoffman2014no}
{\sc M.~D. Hoffman and A.~Gelman}, {\em The no-u-turn sampler: adaptively
  setting path lengths in hamiltonian monte carlo.}, Journal of Machine
  Learning Research, 15 (2014), pp.~1593--1623.

\bibitem{hosmer2013applied}
{\sc D.~W. Hosmer~Jr, S.~Lemeshow, and R.~X. Sturdivant}, {\em Applied logistic
  regression}, vol.~398, John Wiley \& Sons, 2013.

\bibitem{jaakkola1997variational}
{\sc T.~Jaakkola and M.~Jordan}, {\em A variational approach to bayesian
  logistic regression models and their extensions}, in Sixth International
  Workshop on Artificial Intelligence and Statistics, vol.~82, 1997.

\bibitem{mihelich2018maximum}
{\sc M.~Mihelich, B.~Dubrulle, D.~Paillard, Q.~Kral, and D.~Faranda}, {\em
  Maximum kolmogorov-sinai entropy versus minimum mixing time in markov
  chains}, Journal of Statistical Physics, 170 (2018), pp.~62--68.

\bibitem{moller1998log}
{\sc J.~Moller, A.~R. Syversveen, and R.~P. Waagepetersen}, {\em Log gaussian
  cox processes}, Scandinavian journal of statistics, 25 (1998), pp.~451--482.

\bibitem{neal2011mcmc}
{\sc R.~M. Neal}, {\em Mcmc using hamiltonian dynamics}, in Handbook of markov
  chain monte carlo, S.~Brooks, A.~Gelman, G.~Jones, and X.-L. Meng, eds.,
  Chapman \& Hall/CRC, 2011, pp.~131--162.

\bibitem{nutsmatlab}
{\sc A.~Nishimura}, {\em (recycled) no-u-turn-sampler : Matlab implementation},
  2017.

\bibitem{Pasarica2007Adaptively}
{\sc C.~Pasarica and A.~Gelman}, {\em Adaptively scaling the metropolis
  algorithm using expected squared jumped distance}, Statistica Sinica, 20
  (2007), pp.~343--364.

\bibitem{sinai2009kolmogorov}
{\sc Y.~Sinai}, {\em Kolmogorov-sinai entropy}, Scholarpedia, 4 (2009),
  p.~2034.

\bibitem{strathmann2015gradient}
{\sc H.~Strathmann, D.~Sejdinovic, S.~Livingstone, Z.~Szabo, and A.~Gretton},
  {\em Gradient-free hamiltonian monte carlo with efficient kernel exponential
  families}, in Advances in Neural Information Processing Systems, 2015,
  pp.~955--963.

\bibitem{teng2017bayesian}
{\sc M.~Teng, F.~Nathoo, and T.~D. Johnson}, {\em Bayesian computation for
  log-gaussian cox processes: a comparative analysis of methods}, Journal of
  statistical computation and simulation, 87 (2017), pp.~2227--2252.

\bibitem{wang2013adaptive}
{\sc Z.~Wang, S.~Mohamed, and N.~Freitas}, {\em Adaptive hamiltonian and
  riemann manifold monte carlo}, in International Conference on Machine
  Learning, 2013, pp.~1462--1470.

\bibitem{mcesmatlab}
{\sc T.~Yu}, {\em Maximum conditional entropy sampler: Matlab implementation}.
\newblock {https://github.com/SiriusYtc/MCES}, 2017.

\bibitem{zhang2011quasi}
{\sc Y.~Zhang and C.~A. Sutton}, {\em Quasi-newton methods for markov chain
  monte carlo}, in Advances in Neural Information Processing Systems, 2011,
  pp.~2393--2401.

\end{thebibliography}

\end{document}